\documentclass[aps,prd,amsmath,amssymb,twocolumn,preprintnumbers]{revtex4}
\pdfoutput=1
\usepackage{graphicx}
\usepackage{color}

\newcommand{\bmp}{\noindent\begin{minipage}{16cm}}
\newcommand{\emp}{\end{minipage}\vskip 7mm} 

\usepackage{bm}
\usepackage{bbm}

\newcommand{\beq}{\begin{equation}}
\newcommand{\eeq}{\end{equation}}
\newcommand{\bea}{\begin{eqnarray}}
\newcommand{\eea}{\end{eqnarray}}
\newcommand{\ba}{\begin{array}}
\newcommand{\ea}{\end{array}}
\newcommand{\bi}{\begin{itemize}}
\newcommand{\ei}{\end{itemize}}
\newcommand{\bn}{\begin{enumerate}}
\newcommand{\en}{\end{enumerate}}
\newcommand{\bc}{\begin{center}}
\newcommand{\ec}{\end{center}}

\newcommand{\gsim}{\lower.7ex\hbox{$\;\stackrel{\textstyle>}{\sim}\;$}}
\newcommand{\lsim}{\lower.7ex\hbox{$\;\stackrel{\textstyle<}{\sim}\;$}}

\definecolor{rossoCP3}{cmyk}{0,.88,.77,.40}

\begin{document}
\title{\Large  \color{rossoCP3} Cosmic-ray Sum Rules} 
\author{Mads T. {\sc Frandsen}$^{\color{rossoCP3}{\spadesuit}}$}
\email{m.frandsen1@physics.ox.ac.uk} 
\author{Isabella {\sc  Masina}$^{\color{rossoCP3}{\clubsuit},{\heartsuit}}$}
\email{masina@fe.infn.it} 
\author{Francesco {\sc Sannino}$^{\color{rossoCP3}{\heartsuit}}$}
\email{sannino@cp3.sdu.dk} 

\affiliation{
$^{\color{rossoCP3}{\spadesuit}}${Rudolf Peierls Centre for Theoretical Physics, University of Oxford, 1 Keble Road, Oxford OX1 3NP, United Kingdom}}
\affiliation{
$^{\color{rossoCP3}{\clubsuit}}${Dip.~di Fisica dell'Universit\`a di Ferrara and INFN Sez.~di Ferrara, Via Saragat 1, I-44100 Ferrara, Italy}}
\affiliation{
$^{\color{rossoCP3}{\heartsuit}}${ CP}$^{ \bf 3}${-Origins}, 
University of Southern Denmark,  Campusvej 55, DK-5230 Odense M, Denmark.}
\begin{abstract}
 We introduce new sum rules allowing to determine universal properties of the unknown component of the cosmic rays and show how {they} can be used to predict the positron fraction at energies not yet explored by current experiments and to constrain specific models.  
  \\[.1cm]
{\footnotesize  \it Preprint: CP$^3$-Origins-2010-48}
\end{abstract}

\maketitle
Shedding light on astrophysical and particle physics origins of cosmic-rays can lead to breakthroughs in our understanding of the fundamental laws ruling the universe. The goal of this paper is to introduce a new model independent approach for combining observations from different cosmic-rays experiments.  We show that our results lead to predictions relevant for future observations and to constraints useful to guide model builders. The data recently collected by PAMELA \cite{Adriani:2008zr} indicate that there is a positron  
excess in the cosmic ray (CR) energy spectrum above $10$ GeV.
 {The} rising behavior {observed by PAMELA} does not fit previous estimates of the CR formation and propagation implying the possible existence of 
a direct excess of CR {positrons} of unknown origins. 
Interestingly PAMELA's data show a clear feature of {such a positron excess but no excess in the anti-protons}. {While ATIC \cite{:2008zzr} and PPB-BETS \cite{Yoshida:2008zzc} reported unexpected
structure in the all-electron spectrum in the range $100$~GeV- $1$~TeV, the
picture has changed with the higher-statistics measurements by
FERMI-LAT \cite{Abdo:2009zk} and HESS \cite{Aharonian:2008aa}, leading to a possible slight additional
unknown component in the CR $e^{\pm}$ flux over and above the specific Moskalenko and Strong model
prediction \cite{Strong:1998pw, Baltz:1998xv} which assumes a single-power-law injection spectrum of $e^{\pm}$
cosmic rays.} These interesting features have drawn much attention, and many explanations have been proposed: {} For example,
these excesses could be due to an inadequate account of the CR background 
in previous modeling; The presence of new astrophysical sources;
They could also originate from annihilations and/or decays of dark matter. 
We refer to  \cite{Fan:2010yq} for a recent review.  Whatever the origin of these excesses might be, we show that we can derive novel constraints able to shed light 
on the physical nature of their source and/or propagation. We start by writing the observed flux of electrons and positrons as the sum of two contributions:
A background component, $\phi_{\pm}^B$, due to known astrophysical sources (at least for the electrons), 
and an unknown component, $\phi_{\pm}^U$, in formulae:
\beq
\phi_\pm = \phi_\pm^U + \phi_\pm^B  \ .
\eeq
The component $\phi_{\pm}^U$ is the one needed to explain the features in the spectra 
observed by PAMELA and FERMI-LAT. These experiments measure respectively the positron fraction and the total electron and positron fluxes 
as a function of the energy $E$ of the detected $e^\pm$, i.e.: 
\beq
P(E) = \frac{\phi_+(E)}{\phi_+(E) + \phi_-(E)}\ , \qquad  F(E) = \phi_+(E) + \phi_-(E) \ .
\eeq
The left-hand side of the equations above refer to the experimental measures. 
The contribution from the unknown source is then: 
\begin{eqnarray}
\phi_+^U(E) & = &  P(E)~F(E) -\phi_+^B(E) \ ,  \\ 
\phi_-^U(E) & = & F(E)~ \left(1-P(E)\right) -\phi_-^B(E) \ .
\end{eqnarray}
In terms of their difference and sum:
\begin{eqnarray}
\phi_+^U(E) -\phi_-^U(E) &=&   F(E)~(2 P(E)-1) +(\phi_-^B(E) -\phi_+^B(E) ) \ ,\nonumber \\  
\phi_+^U(E) +\phi_-^U(E) &=&  F(E)- (\phi_-^B(E)+\phi_+^B(E)) \nonumber \ . \\
&& 
\label{sumrules}
\end{eqnarray}
The latter equation implies $F(E)\geq \phi_-^B(E)+\phi_+^B(E)$.
We model the background spectrum using $\phi_\pm^B(E)=N_B B^\pm(E)$, where
$N_{B}$ is a normalization coefficient such that $F(E)/ (B^-(E)+ B^+(E))\geq N_B$
and $B^\pm(E)$ are provided using specific astrophysical models. 
In this paper we adopt the popular Moskalenko and Strong model \cite{Strong:1998pw, Baltz:1998xv},
for which $N_B$ is less than  $0.75$  and $B^\pm(E)$ are given by: 
\bea
B^+&=&  \frac{ 4.5 E^{0.7}}{1 + 650E^{2.3} + 1500E^{4.2}} \ ,
 \\
B^-&=& B_1^- + B_2^- \ ,\\ 
B_1^-&=& \frac{ 0.16 E^{-1.1}} {1 + 11 E^{0.9} + 3.2 E^{2.15}}  \ , 
  \\
B_2^-&=& \frac{ 0.70 E^{0.7}}{1 + 110 E^{1.5} + 600 E^{2.9} + 580 E^{4.2}} \ ,
\eea
where $E$ is measured in ${\rm GeV}$ and the $B$s in ${\rm GeV}^{-1} {\rm cm}^{-2}{\rm sec}^{-1}{\rm sr}^{-1}$ units. We checked that our results remain unchanged when replacing the parameterization above with the one adopted by the Fermi Collaboration (model zero) \cite{Grasso:2009ma, Ibarra:2009dr}. 

It is convenient to introduce the following parameter: 
\beq
r_U(E) \equiv \frac{\phi_-^U(E)}{\phi_+^U(E)}= \frac{F(E) ~(1-P(E))-\phi_-^B(E)}{P(E)~F(E)-\phi_+^B(E)}~~.
\label{ruu}
\eeq
This equation can be rewritten as
\beq
R(E) \equiv \frac{F(E)}{B^-(E)} ~\frac{ 1-(1+r_U(E)) P(E)}{1-r_U(E) \frac{\phi_+^B(E)}{\phi_-^B(E)}} = N_B \ .
\label{sumrulefinal}
\eeq
Although the sum rule $R(E)$ seems to depend on the energy it should, in fact, be a constant as is clear from the right hand side of the previous equation.  
{\it This leads to a nontrivial constraint linking together in an explicit form the experimental results, the model of the backgrounds and the dependence on the energy of the unknown components.}  

\begin{figure}[th!]
\begin{center} 
\includegraphics[width=5.5cm]{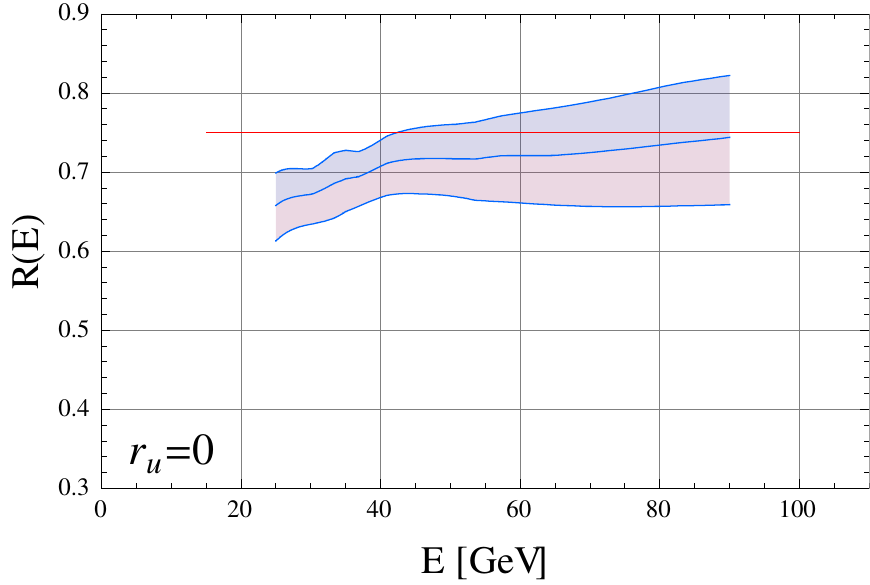} \includegraphics[width=5.5cm]{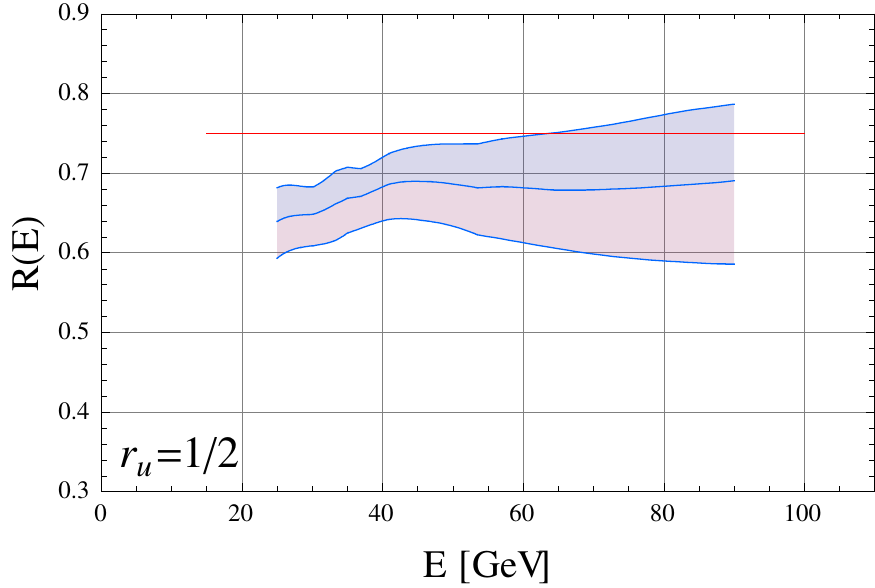} 
\includegraphics[width=5.5cm]{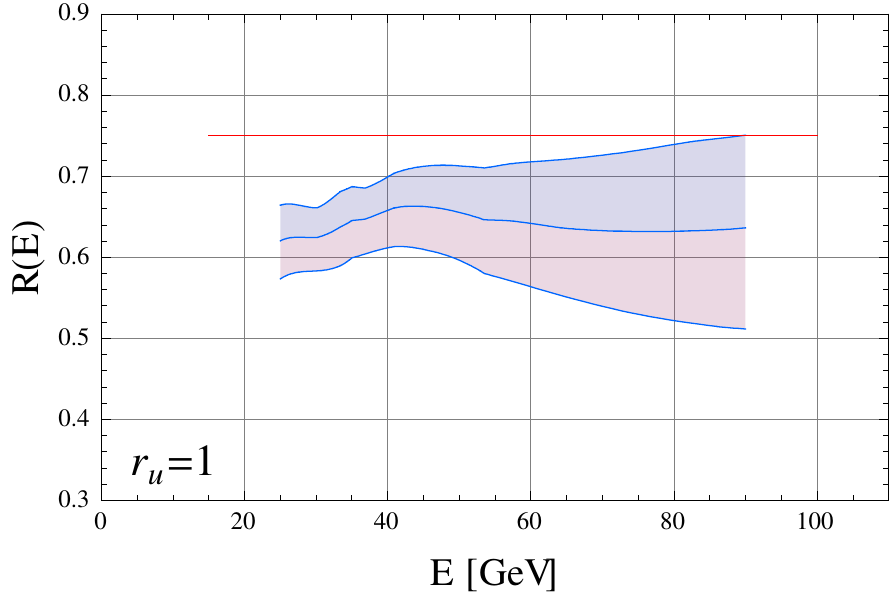} \includegraphics[width=5.5cm]{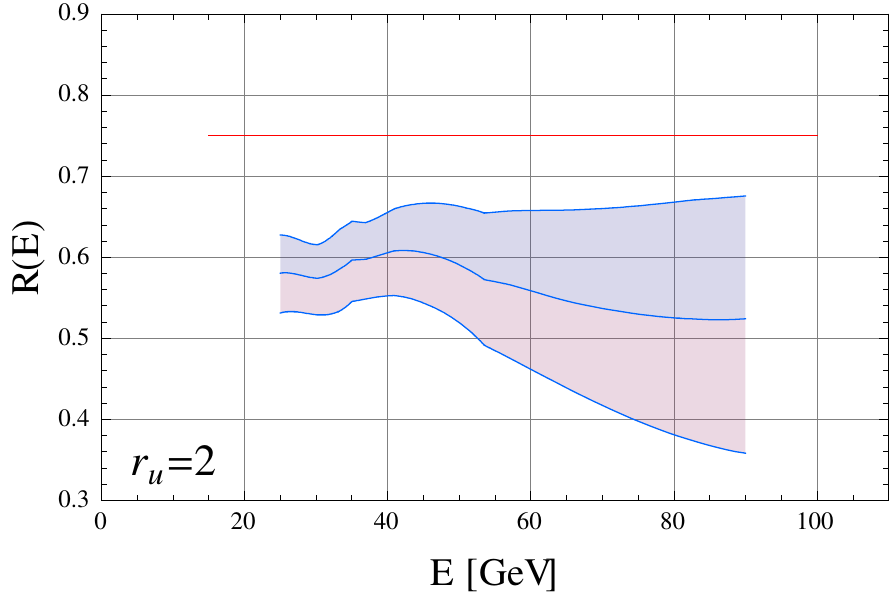} 
\end{center}
\caption{Ratio $R(E)$ as a function of the energy $E$ of electrons and positrons
and for vales of $r_U=0,1/2,1,2$, from top to bottom.
The shaded region accounts for the one sigma error in PAMELA and FERMI-LAT. Secondaries are estimated according to 
the expressions in \cite{Strong:1998pw, Baltz:1998xv}.}
\label{Ratios}
\end{figure}

\begin{figure}[h!]
\begin{center} 
\includegraphics[width=6cm]{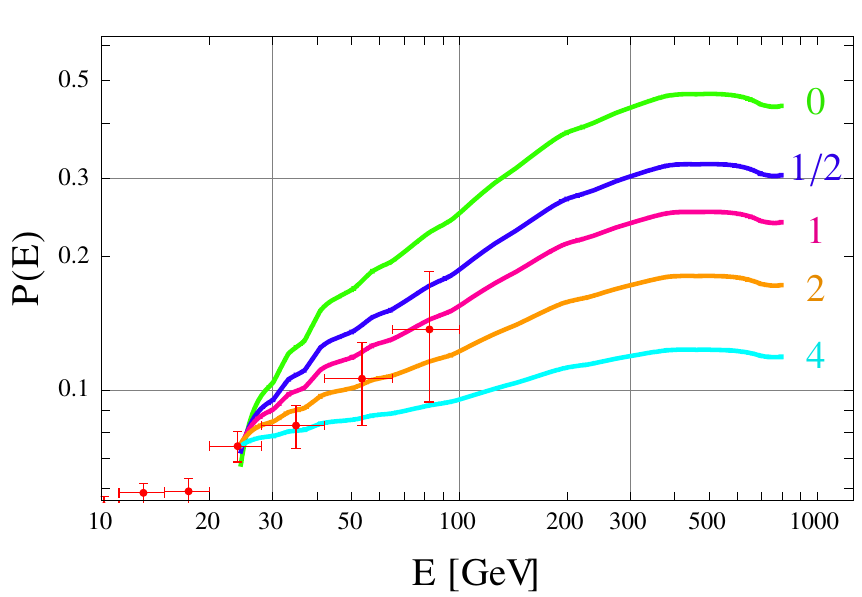} 
\end{center}
\caption{Model independent prediction for the positron fraction $P(E)$ as a function of the energy $E$ 
of electrons and positrons
and for vales of $r_U=0,1/2,1,2,4$, from top to bottom.
Secondaries are estimated according to 
the expressions in \cite{Strong:1998pw, Baltz:1998xv} reported in the main text and the derived $N_B$ values:
$N_B = 0.66, 0.64, 0.62, 0.58, 0.50$.}
\label{predictionP}
\end{figure}

We now turn to the actual data and show in which way the sum rule \eqref{sumrulefinal} provides essential information on the unknown components of the CRs. Since we use simultaneously the results of FERMI-LAT and PAMELA we are obliged to consider only the common 
energy range.  Note that the CRs energy range below $10-20~{\rm GeVs}$, where the spectrum 
is affected by the Sun, is outside the common range. 
Within the relevant  but limited range of energies we will consider here it is therefore sensible 
to assume $r_U$ to be nearly constant. We find useful to plot the function $R(E)$ for different values of $r_U$  in order to test the  sum rule \eqref{sumrulefinal}. This would imply that this function is independent of the energy. 
The associated constant value would then be identified with the background CRs normalization factor $N_B$. 
We report the results in Fig.~\ref{Ratios}. The straight (red) line is the $N_B = 0.75$ value which is 
the largest one can assume for the background not to be larger than the FERMI-LAT results. 
We observe that there is a clear tendency for the combined data to predict a lower value of 
the constant $N_B = 0.66, 0.64, 0.62, 0.58$ for increasing value of the ratio $r_U = 0,1/2,1,2$. 
This is clear when looking, from top to bottom, at the different panels of Fig.~\ref{Ratios}. 
It is interesting to note that we find a plateau, in the relevant energy range, up to $r_U$ near 
the value of $2$ when $R(E)$ starts showing some deviation. The PAMELA and FERMI-LAT mean data have been used to determine the mean values of $R(E)$ in the different panels of Fig.~\ref{Ratios}. As explained above we compare the data only in the energy range where the two experiments overlap. In the various panels we also show the uncertainties around the mean value which we have determined in the following way. We maximized  (minimized) $R(E)$ in \eqref{sumrulefinal} by using the one sigma deviations coming from both PAMELA and FERMI-LAT. This resulted in the shaded band around the mean value.  Given the large uncertainties we cannot 
yet provide a more solid conclusion, however we can use the derived normalization for each different 
ratios of the unknown components to {\it predict} the positron fraction at energies higher than 
the ones provided so far by PAMELA.
In order to be able to make such a prediction we first rewrite \eqref{ruu} as follows:
\beq
P(E) = \frac{1}{1+r_U} \left(1-\frac{\phi^B_-(E)}{F(E)}(1-r_U \frac{\phi^B_+(E)}{\phi^B_-(E)})\right) \ , 
\eeq
where we use for each $r_U$ the estimated associated $N_B$. The different predictions for the positron 
fraction, assuming that $r_U$ remains constant over the entire energy range, 
up to $1000$~GeV are shown in Fig.~\ref{predictionP}.
The resulting picture disfavors both very small and very large values of $r_U$ which, in turn, means that one expects 
{the electron fraction to neither be too small nor to indicate a large proportion of positrons}.  

Using the sum rule will allow to extract vital information from the data as they become more and more accurate. 
\begin{figure}[hbt!]
\begin{center} 
\includegraphics[width=6cm]{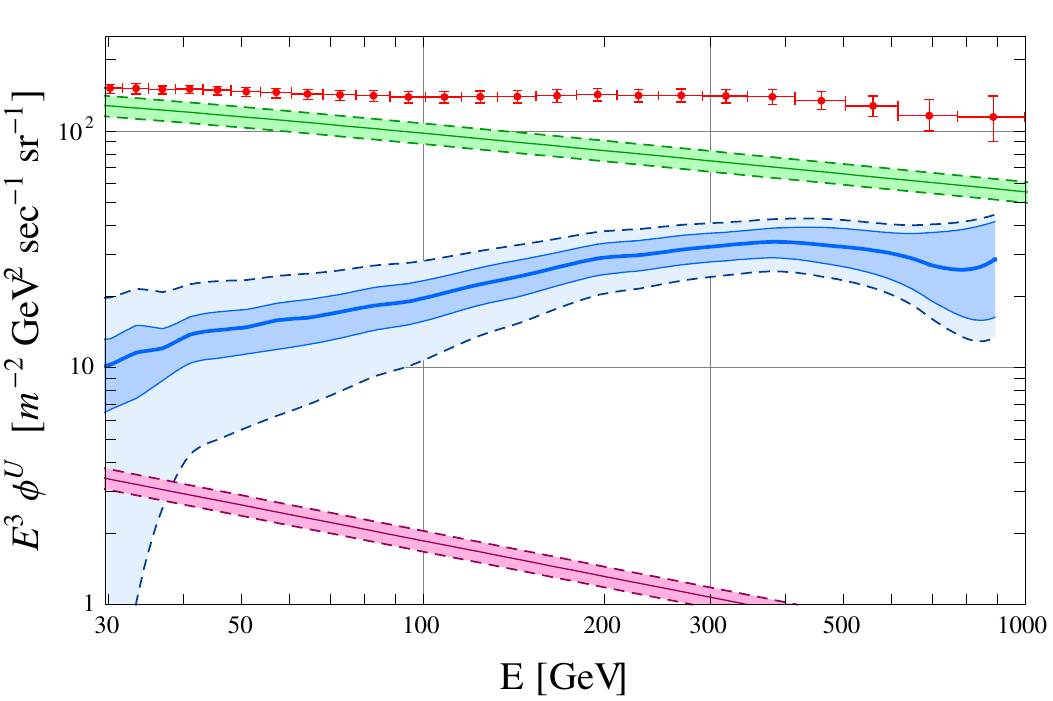}\end{center}
\caption{We display $\phi^{U}(E) = \phi_\pm^U(E)$  corresponding to the central shaded region  
for the whole energy range of the FERMI-LAT experiment.  
The upper red points are the actual FERMI-LAT data,
and the background of electrons is the upper green stripe while the positrons is the lower magenta stripe. See the text for further explanation. }
\label{Sum}
\end{figure}
The special case $r_U=1$ is an automatic prediction of a great deal of models for dark matter which assume that charge symmetry holds both in the production and propagation of the unknown component of the CRs. 

It is, therefore, clear that our formalism is more universal given that we have made {\it no assumption}  
in the derivation of the sum rules above. It is however, useful {to adopt the model assumption of $r_U=1$} 
to derive the further constraint: 
\begin{equation}
\phi_\pm^U(E)  =    \frac{ F(E)-(\phi^B_-(E) +\phi^B_+(E) )}{2} \ ,\label{SR2}
\end{equation}
for which we now use $N_B = 0.62$. {We show in   Fig.~\ref{Sum}  the sensitivity of the unknown 
flux $\phi^{U}(E) = \phi_\pm^U(E)$ on the experimental errors as well as  on  a 10\% arbitrary 
variation of the background spectrum.  $\phi^{U}(E)$ is the difference between the FERMI-LAT data 
(the red dotted points) and the  positron (the magenta lower line) and electron (upper green line) 
backgrounds. The vertical spread associated with the inner (blue) darkest region of $\phi^{U}(E)$  is  due to FERMI-LAT error bars while the outer lighter shaded region also includes a 10\% variation of the background. This does not imply that the backgrounds are known to a 10\% accuracy but merely shows how sensitive $\phi^{U}(E)$ can be to such a variation.  Although we investigated the case $r_u=1$ it is clear that our sum rules can be used to test any model of the unknown component. }

The sum rules introduced here are general and can be extended also to the proton and antiproton fraction. 
We have shown that current data still allow for approximately equal contributions of the electrons and positrons from the unknown components of the associated CRs, but disfavor electron to positron fractions smaller than one half and larger than four. 

The current model independent analysis of the combined PAMELA and FERMI-LAT data shows that we will be able to deduce, thanks to the new constraints, vital information on the nature of the source and the propagation of the CRs. We have also demonstrated that the typical oversimplifying model assumption used so far constitutes a small portion of the allowed models still left unconstrained by the present data. Finally our sum rules can be easily used to constrain any specific model while we were able to predict, in a model independent way, the positron fraction at energies higher than the ones explored so far.





\end{document}